\documentclass[12pt, a4paper, twoside]{article}
 \usepackage[font=small,format=plain,labelfont=bf,up,textfont=normal,up,justification=justified,singlelinecheck=false]{caption}

\usepackage[a4paper, left=3cm,right=2cm]{geometry}
\usepackage{hyperref}
\usepackage{amsfonts}
\usepackage{amsmath}
\usepackage{setspace}
\usepackage{color}

\usepackage[utf8,applemac]{inputenc}
\usepackage{tensor}
\usepackage{cite}
\usepackage{tikz}
\usepackage{graphicx}
\usepackage{dcolumn}
\usepackage{bm}
\usepackage[justification=centering]{caption} 

\newcommand{\bea}{\begin{eqnarray}}
\newcommand{\eea}{\end{eqnarray}}
\newcommand{\be}{\begin{equation}}
\newcommand{\ee}{\end{equation}}
\def\nn{\nonumber}

  \newcommand{\beqs}{\begin{eqnarray}}
\newcommand{\eeqs}{\end{eqnarray}}

\title{On co-dimension two defect operators}

\begin{document}
\setcounter{tocdepth}{2}
\begin{titlepage}
\begin{flushright}\vspace{-3cm}
{\small
\today }\end{flushright}
\vspace{0.5cm}

\begin{center}
{{ \LARGE{\bf{On co-dimension two defect operators\ }}}} \vspace{5mm}

\centerline{\large{\bf{Jiang Long\footnote{e-mail:
Jiang.Long@ulb.ac.be}}}}

\vspace{2mm}
\normalsize
\bigskip\medskip
\textit{Universit\'{e} Libre de Bruxelles and International Solvay Institutes\\
CP 231, B-1050 Brussels, Belgium
}

\vspace{25mm}

\begin{abstract}
\noindent
{Conformal symmetry is broken by a flat or spherical defect operator $\mathcal{D}$. We show that this defect operator,  may be identified as a pair of twist operators which are inserted at the tips of its causal diamond.  Any $k-$point correlation function in a flat or spherical defect CFT is equivalent to a $(k+2)-$point correlation function.  We reproduce one point correlation functions and also solve two point correlation functions in defect CFTs .  Mutual R\'enyi entropy is computed and agrees with previous result in a certain limit. We conjecture there may be universal terms in general co-dimension two defect CFTs.}
\end{abstract}


\end{center}

\end{titlepage}

\newpage
\tableofcontents

\section{Introduction}
All dynamical information of a conformal field theory are encoded in the coefficients of three point functions . The form of two and three point correlation functions are fixed by conformal symmetry. Higher point functions can be reduced to lower point functions by operator product expansion.  The operator product expansion can be done in different channels, leading to crossing symmetry in four point correlation functions. 

One can break  conformal symmetry by inserting an extended defect operator $\mathcal{D}$\cite{Diehl:1981,Cardy:1984, Cardy:1989, Cardy:1991,McAvity:9302, McAvity:9505}, see recent progress on defect CFTs\cite{Liendo:1210,Billo:1304,Gaiotto:1310,Gliozzi:1502, Jensen:1509,Bianchi:1511,Marco:1601, Abhijit:1602, Gliozzi:1605,Liendo:1608}.  A co-dimension $q$ flat or spherical defect operator is special as it preserves $SO(p+1,1)\times SO(q-1,1)$ symmetry\footnote{We will study Lorentz CFT in this work.} where $p+q=d$. In this case, one point function of any primary operator  is completely fixed up to a constant  while two point function of primary operators is constrained to be a summation of conformal partial waves.

Among all the flat or spherical defect operators
, the co-dimension two operator ($q=2$) is especially interesting. It may have applications to R\'enyi entropy(and entanglement entropy)\cite{Srednicki:9003, Holzhey:9403,Ryu:0603,Ryu:0605 }. R\'enyi entropy is actually a partition function on  a $n$ replicated manifold. One divides the space into a region $A$ and its complement $A^c$ and glue different replicas  along the boundary of $A$. Alternatively, one may introduce a co-dimension two defect operator to implement the replica trick. In two dimensions, the defect operator is a pair of twist operators which are inserted at the boundary of the interval $A$\cite{Cardy:0405, Cardy:0505}. In higher dimensions, the defect operator may be a surface operator\cite{Kapustin:0501,Hung:1110,Hung:1407}. 

\begin{figure}
\centering
 \includegraphics[width=0.5\textwidth]{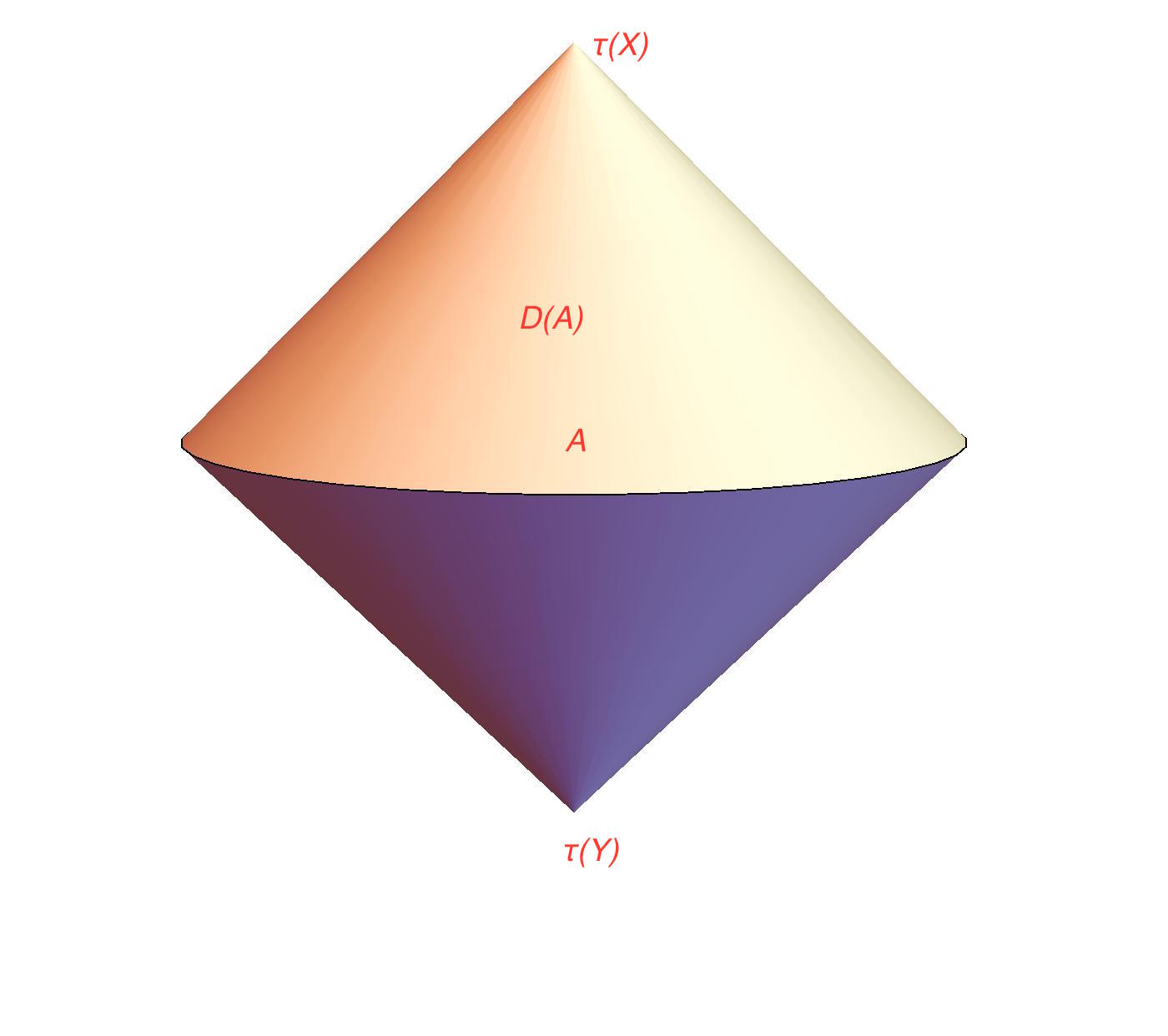}  
     \caption{Causal diamond of a spherical defect.}\label{causal}
\end{figure}

 However, we will see that the flat or spherical defect operator may be identified as a pair of operators
\be
\mathcal{D}=\tau(X)\tau(Y).\label{defect}
\ee
We will call  $\tau$  a twist operator in reference to entanglement entropy. 
The two twist operators are inserted at the tips of the causal diamond of $A$. 
Figure \ref{causal} is an illustration of a spherical defect operator and its associated causal diamond. 
Suppose there is a $d-1$ dimensional ball $A$ at constant time on flat spacetime. The spherical defect operator sits at the boundary of $A$ which is a $d-2$ sphere. The defect operator $\mathcal{D}$ divides the space into $A$ and its complement $A^c$.  There is a causal diamond $D(A)$ of region $A$. The two twist operators sit at the tips of the causal diamond $D(A)$, in the figure, we denote them as point X and Y. One is in the future and the other one is in the past.
This is the main geometric picture of a spherical defect operator. 

(\ref{defect}) is an operator identity.  Any $k$-point correlation function in a defect CFT will be a $(k+2)$-point correlation function in the vacuum, 
\be
<\mathcal{O}_1\cdots\mathcal{O}_k>_{\mathcal{D}}=\frac{<\mathcal{O}_1\cdots\mathcal{O}_k\tau(X)\tau(Y)>}{<\tau(X)\tau(Y)>}\label{kk2}.
\ee
When $k=1$, the form on the left hand side of (\ref{kk2}) is fixed by symmetry\cite{Marco:1601},  the right hand side can also be fixed by conformal symmetry as it is a three point function. we checked that they lead to the same answer once we properly identify the coefficients in front of them. When $k=2$, the left hand side can be expanded into conformal partial waves which satisfy special equation of motions, the right hand side can also be expanded into conformal partial waves whose form can be fixed by the conformal Casimir equation\cite{Dolan:0011, Dolan:0309}. We checked that they can be mapped to each other after properly identify the conformal partial waves.  For higher point correlation functions($k>2$), we can always use operator product expansion to reduces them to lower point correlation functions. Hence, (\ref{kk2}) will be valid for arbitrary correlation functions. 

When there are multi-defects in the CFT,  the correlation function should be \footnote{We assume they are disjoint.}
\be
<\mathcal{O}_1\cdots\mathcal{O}_k>_{\mathcal{D}}=\frac{<\mathcal{O}_1\cdots\mathcal{O}_k\prod_{i=1}^m\tau(X_i)\tau(Y_i)>}{<\prod_{i=1}^m\tau(X_i)\tau(Y_i)>}\label{kkm}.
\ee
We apply (\ref{kkm}) to compute mutual R\'enyi entropy of two spherical region and find perfect agreement with previous paper\cite{Cardy:1304, Dong:1601}.

This paper is organized as follows. Section two is to review main results on flat or spherical defect operators which are fixed by symmetry. We will check the identity (\ref{defect}) in section three. We will study operator product expansion of defect operators $\mathcal{D}$ in section four.  In section five, we will apply our formula to compute mutual R\'enyi entropy of a free theory and holographic CFTs.  We will end with discussions and point out some future directions. Conventions and technical details are collected in the Appendices.

\section{Defect CFT}
This section is to collect useful results in defect CFTs\cite{Marco:1601}. 
A defect CFT is a CFT with defect operator $\mathcal{D}$ being inserted.  Correlation functions of primary operators are defined as
\be
<\mathcal{O}_1(x_1)\cdots\mathcal{O}_k(x_k)>_{\mathcal{D}}=\frac{<\mathcal{O}_1(x_1)\cdots\mathcal{O}_k(x_k)\mathcal{D}>}{<\mathcal{D}>}. 
\ee
We will study flat or spherical defect CFTs.  As we can always transform a flat defect  to a spherical defect by conformal transformation, it is sufficient to show the results of flat defect.  We assume the defect sits at $t=0,x^1=0$. 
A primary operator $\mathcal{O}$ sits at $x=(x^a, y^i)$ where $a=0,1$ label the transverse directions of the defect and $i=2,\cdots, d-1$ label the parallel directions of the defect. 
The one point function of a primary operator can be fixed up to a constant,  for example, a scalar primary operator 
\be
<\mathcal{O}(x)>_{\mathcal{D}}=\frac{a_{\Delta,0}}{|x|^{\Delta}}. \label{scalar}
\ee
$\Delta$ is the conformal dimension of $\mathcal{O}$.  $|x|$ is the norm of the transverse direction. The first subindex of $a_{\Delta,J}$ is the conformal dimension of operator $\mathcal{O}$, the second subindex is the spin of operator $\mathcal{O}$. For a spin two primary operator, its one point function is 
\bea
<\mathcal{O}_{ab}(x)>_{\mathcal{D}}&=&-a_{\Delta,2}\frac{(d-1)\eta_{ab}-d n_an_b}{|x|^{\Delta}},\nn\\
<\mathcal{O}_{ai}(x)>_{\mathcal{D}}&=&0,\label{spin2}\\
<\mathcal{O}_{ij}(x)>_{\mathcal{D}}&=&a_{\Delta,2}\frac{\eta_{ij}}{|x|^{\Delta}},\nn
\eea
where $n^a$ is an unit vector $n^a=\frac{x^a}{|x|}$.  One point function of any even higher spin operators is also fixed up to a coefficient $a_{\Delta,J}$.  It would be convenient to use embedding formalism to organise the result.  In Appendix B we introduce this formalism and collect necessary results for interested reader. 
Curiously, for odd spin operator, one must introduce parity violating terms.  For spin 1, the answer is 
\be
<\mathcal{O}_{a}>_{\mathcal{D}}=a_{\Delta,1}\frac{\epsilon_{ab}x^b}{|x|^{\Delta+1}}, \ <\mathcal{O}_{i}>_{\mathcal{D}}=0
\ee
where we introduced a parity violating $\epsilon$ tensor in orthogonal direction, $\epsilon_{01}=1$.  

 Two point function of scalar primary operator can be written as the cross ratio $\xi,\phi$ as 
\be
<\mathcal{O}_1(x_1)\mathcal{O}_2(x_2)>_{\mathcal{D}}=\frac{f_{12}(\xi,\phi)}{\xi^{\frac{\Delta_1+\Delta_2}{2}}|x_1|^{\Delta_1}|x_2|^{\Delta_2}},
\ee
where 
\be
\xi=\frac{x_{12}^2+y_{12}^2}{|x_1||x_2|},\ \cos\phi=\frac{x_1^ax_2^b\eta_{ab}}{|x_1||x_2|}
\ee
with
\be
x_{12}^2=\eta_{ab}(x_1^a-x_2^a)(x_1^b-x_2^b),\ y_{12}^2=\eta_{ij}(y_1^i-y_2^i)(y_1^j-y_2^j).
\ee
The function $f_{12}(\xi,\phi)$ can be expanded in terms of CFT conformal partial waves, 
\be
f_{12}(\xi,\phi)=\sum_{\mathcal{O}_{\Delta,J}}c_{12\mathcal{O}_{\Delta,J}}a_{\mathcal{O}_{\Delta,J}}F_{\Delta,J}(\xi,\phi),
\ee
where $c_{12\mathcal{O}_{\Delta,J}}$ is three point function coefficient of $<\mathcal{O}_1\mathcal{O}_2\mathcal{O}_{\Delta,J}>$ and $a_{\mathcal{O}_{\Delta,J}}$ is the coefficient of one point function coefficient of $\mathcal{O}_{\Delta,J}$ in defect CFT. The conformal block $F_{\Delta,J}(\xi,\phi)$ satisfy a second order differential equation \cite{Marco:1601}
\be
\mathcal{D}_{bulk}F_{\Delta,J}(\xi,\phi)=0.\label{ca1}
\ee
The explicit form of the differential operator $\mathcal{D}_{bulk}$ are collected in Appendix C.

\section{Defect operator $\mathcal{D}$}
In this section, we will study the defect operator $\mathcal{D}$.  As we have mentioned briefly in the introduction, we expect $\mathcal{D}$ to be equivalent to a pair of twist operators 
\be
\mathcal{D}=\tau(X)\tau(Y)\label{defecttwist}
\ee
For a flat defect which is placed at $t=0,x^1=0$, its associated causal diamond is $D(x<0)$\footnote{There is another causal diamond which is $D(x>0)$.  As we are trying to study correlation functions with operator inserted at $x>0$, we use causal diamond $D(x<0)$ to avoid possible singularity.}. The twist operators are placed at\footnote{The parallel coordinates are irrelevant as we will set $T\to\infty$ finally.} $(T,-T)$ and $(-T,-T)$ with $T\to\infty$.  The scaling dimension of the twist operator is $\delta$.
\begin{figure}
\centering
 \includegraphics[width=0.5\textwidth]{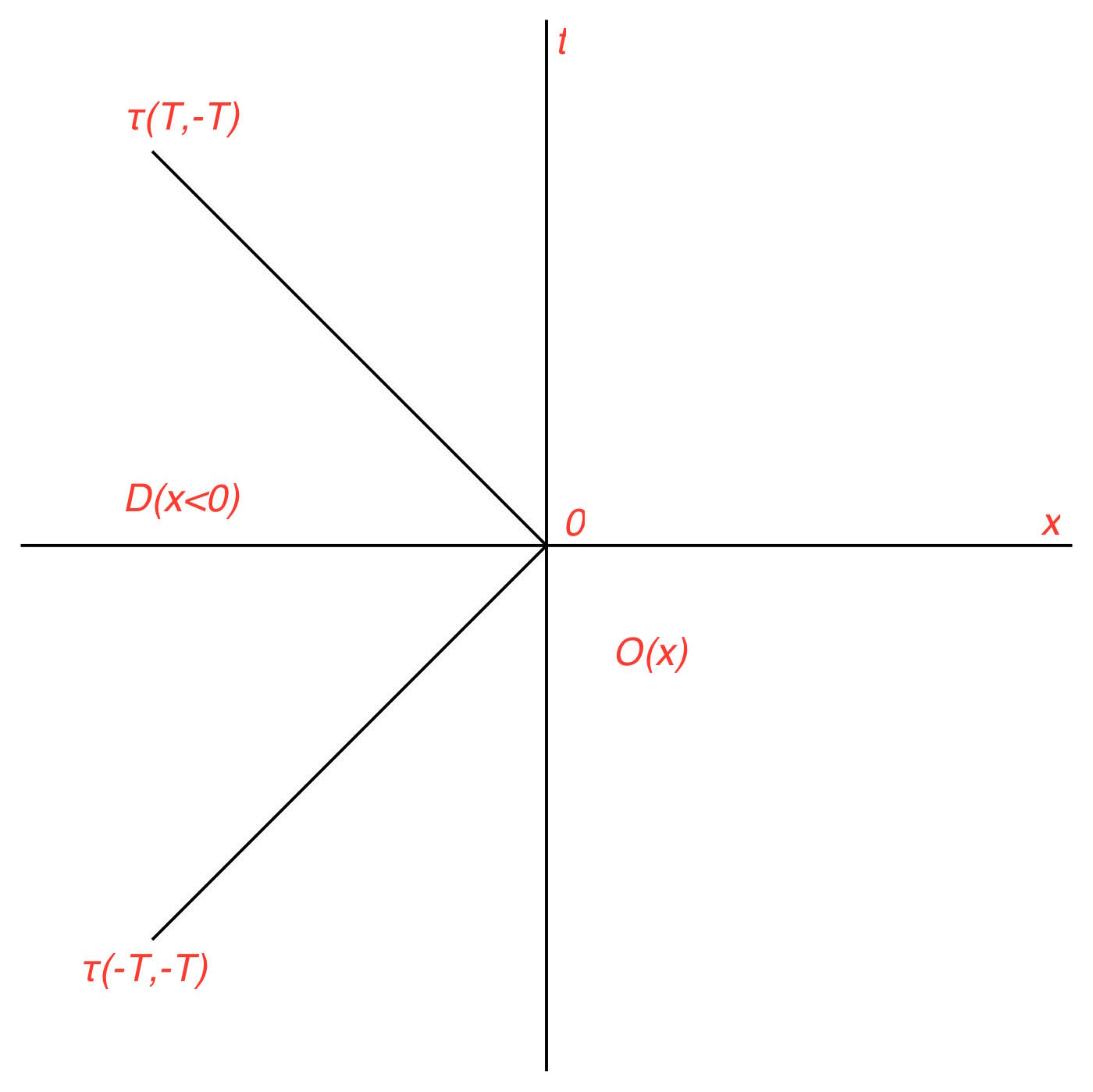}  
     \caption{Flat defect sits at $x=0$. Two twist operators are inserted at $(T,-T)$ and $(-T,-T)$ with $T\to\infty$.}\label{causalflat}
\end{figure}
\subsection{One point functions}
For a primary scalar operator $\mathcal{O}$ with dimension $\Delta$, its one point function is 
\be
<\mathcal{O}(x)>_{\mathcal{D}}=\frac{<\mathcal{O}(x)\mathcal{D}>}{<\mathcal{D}>}=\lim_{T\to\infty}\frac{<\mathcal{O}(x)\tau(T,-T)\tau(-T,-T)>}{<\tau(T,-T)\tau(-T,-T)>}=\frac{c_{\mathcal{O}\tau\tau}}{|x|^{\Delta}}.\label{scalartwist}
\ee
At the last step, we used the well known fact that three point function is fixed up to a constant. Note (\ref{scalartwist}) is exactly (\ref{scalar}) once we identify
\be
c_{\mathcal{O}\tau\tau}=a_{\Delta,0}.
\ee 
One can also check other spin operators. 

It is curious to understand odd spin result. For $J=1$, our method would imply 
\be
<\mathcal{O}_{\mu}(x)>_{\mathcal{D}}=\lim_{T\to\infty}\frac{<\mathcal{O}_{\mu}(x)\tau(T,-T)\tau(-T,-T)>}{<\tau(T,-T)\tau(-T,-T)>}\label{spin1}
\ee
The three point function of spin 1 vector and two primary operator is 
\be
<\mathcal{O}_{\mu}(x_3)\mathcal{O}(x_1)\mathcal{O}(x_2)>=\frac{c_{12\mathcal{O}_{\mu}}}{x_{12}^{\frac{\Delta_1+\Delta_2-\Delta+1}{2}}x_{23}^{\frac{\Delta_2+\Delta-1-\Delta_3}{2}}x_{31}^{\frac{\Delta-1+\Delta_1-\Delta_2}{2}}}(\frac{x_{31\mu}}{x_{31}^2}-\frac{x_{32\mu}}{x_{32}^2}).
\ee
Substituting it to (\ref{spin1}), we find 
\be
<\mathcal{O}_{a}(x)>_{\mathcal{D}}=a_{\Delta,1}\frac{\epsilon_{ab}x^b}{|x|^{\Delta+1}},\ <\mathcal{O}_{i}(x)>_{\mathcal{D}}=0.
\ee
We have redefined the three point function coefficient of $\mathcal{O}_{\mu}$ and twist operators to be $a_{\Delta,1}$.  Note it matches with previous result. Interestingly, the parity violating term appears naturally. This method can be extended to higher spins. It is more convenient to use embedding formalism in this case.  We list the results below while leaving the derivation to Appendix B. 
\be
<\mathcal{O}_{\Delta,J}(P,Z)>_{\mathcal{D}}=a_{\Delta,J}\frac{(\epsilon_{AB}Z^AP^B)^J}{(P\circ P)^{\frac{\Delta+J}{2}}}.
\ee
It matches known result of spin 0 and spin 1. For spin 2, we use the identity\footnote{Remember we are in Lorenz CFT.}
\be
\epsilon_{AB}\epsilon_{CD}=-\eta_{AB}\eta_{CD}+\eta_{AC}\eta_{BD}.
\ee
Then 
\be
<\mathcal{O}_{\Delta,2}(P,Z)>_{\mathcal{D}}=a_{\Delta,2}\frac{Q_2}{(P\circ P)^{\frac{\Delta}{2}}}, 
\ee
where $Q_J$ is defined as
\be
Q_J=(\frac{(P\circ Z)^2}{P\circ P}-Z\circ Z)^{\frac{J}{2}}. 
\ee

\subsection{Two point functions}
We use scalar operator to illustrate the result. 
\be
<\mathcal{O}_1(x_1)\mathcal{O}_2(x_2)>_{\mathcal{D}}=\lim_{T\to\infty}\frac{<\mathcal{O}_1(x_1)\mathcal{O}_2(x_2)\tau(T,-T)\tau(-T,-T)>}{<\tau(T,-T)\tau(-T,-T)>},\label{4pt}
\ee
Note the four point function on the right hand side can be expanded into conformal partial waves. There are two independent cross ratios,
\be
u=-\frac{x_{12}^2+y_{12}^2}{(x_1^1+x_1^0)(x_2^1-x_2^0)},\ v=\frac{(x_2^1+x_2^0)(x_1^1-x_1^0)}{(x_1^1+x_1^0)(x_2^1-x_2^0)}
\ee
They relate to the two cross ratios defined in the previous section by 
\be
\xi=-\frac{u}{\sqrt{v}},\ \cos\phi=\frac{1+v}{2\sqrt{v}}.\label{cross}
\ee
Then the two point function becomes 
\be
<\mathcal{O}_1(x_1)\mathcal{O}_2(x_2)>_{\mathcal{D}}=(\frac{x_2^1-x_2^0}{x_1^1-x_1^0})^{\frac{\Delta_1-\Delta_2}{2}}\frac{g(u,v)}{(x_{12}^2+y_{12}^2)^{\frac{\Delta_1+\Delta_2}{2}}}.
\ee
$g(u,v)$ is related to $f(\xi,\phi)$ by the following relation
\be
f(\xi,\phi)=g(u,v)v^{-\frac{\Delta_1-\Delta_2}{4}}.\label{fg}
\ee
Note $g(u,v)$ can be expanded in terms of conformal partial waves
\be
g(u,v)=\sum_{\mathcal{O}_{\Delta,J}}c_{12\mathcal{O}_{\Delta,J}}a_{\mathcal{O}_{\Delta,J}}G_{\Delta,J}(u,v),
\ee
where we have defined conformal partial waves as $G_{\Delta,J}$. For example, in $d=4$\cite{Dolan:0309}, 
\bea
G_{\Delta,J}(u,v)&=&(-1)^{J}\frac{z\bar{z}}{z-\bar{z}}(k_{\Delta+J}(z)k_{\Delta-J-2}(\bar{z})-(z\leftrightarrow \bar{z}))\nn\\
k_{a}(z)&=&z^{a}{}_2F_{1}(\frac{a-(\Delta_1-\Delta_2)}{2},\frac{a+(\Delta_3-\Delta_4)}{2},a,z).
\eea
 $z$ and $\bar{z}$ are related to $u,v$ by
\be
u=z\bar{z},v=(1-z)(1-\bar{z}).
\ee
In our case, $\Delta_3=\Delta_4=\delta$ . The conformal partial wave $G_{\Delta,J}$ is the solution of Casimir equation 
\be
D_{CFT}G_{\Delta,J}(u,v)=0.\label{ca2}
\ee
The explicit form of $D_{CFT}$ is collected in Appendix C.  Since $f(\xi,\phi)$ and $g(u,v)$ are related by (\ref{fg}),  (\ref{ca1}) and (\ref{ca2}) should be equivalent to each other by the identification 
\be
F_{\Delta,J}(\xi,\phi)=G_{\Delta,J}(u,v)v^{-\frac{\Delta_1-\Delta_2}{4}}. \label{fg2}
\ee
One can check this identity using the differential operator $\mathcal{D}_{bulk}$ and $D_{CFT}$ in the appendix.  In this way, the solution of  (\ref{ca1}) is (\ref{fg2}).  

In \cite{Marco:1601}, the kinematic part of one point function and two point function (scalar) are determined by analyzing the symmetry constraints on the defect CFT. However, we find the same result by mapping the correlation function in the defect CFT to higher point function in the original CFT. The exact match from the two method should not be an accident, it strongly supports the identification (\ref{defecttwist}). 

\subsection{General correlation functions}
The obvious consequence of $(\ref{defecttwist})$ is to map any $k$-point correlation function of a defect CFT to a $(k+2)$-point correlation function, 
\be
<\mathcal{O}_1(x_1)\cdots\mathcal{O}_k(x_k)>_{\mathcal{D}}=\frac{<\mathcal{O}_1(x_1)\cdots\mathcal{O}_k(x_k)\tau(X)\tau(Y)>}{<\tau(X)\tau(Y)>}.\label{cf}
\ee
The insert positions of the twist operators are at the tips of the causal diamond associated with the defect.   
\subsection{Multi-defects correlation functions}
\begin{figure}
\centering
 \includegraphics[width=0.9\textwidth]{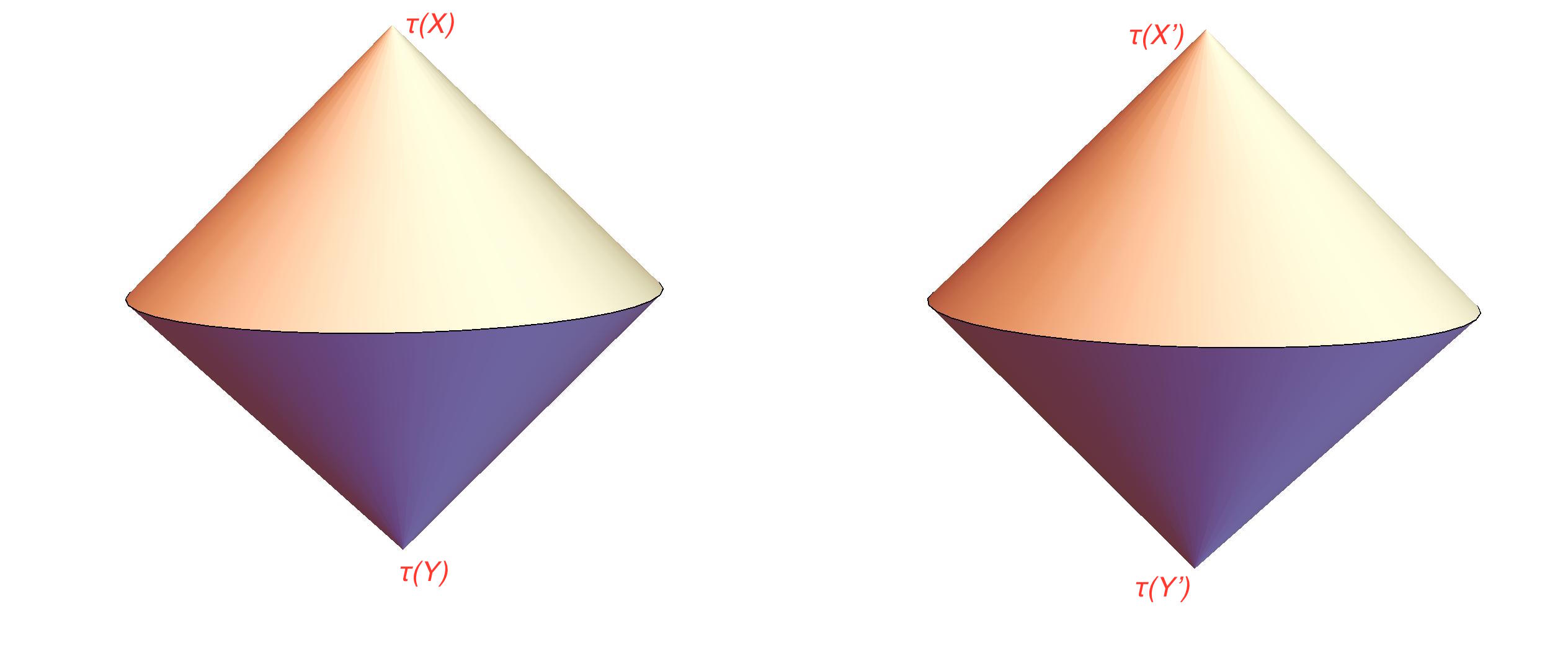}  
     \caption{Two spherical defects. }\label{causaltwos}
\end{figure}

It is sufficient to study two defects case.  We place two spherical defects at\footnote{Note there is only one independent cross ratio in this case,  one can generalize it to  two independent cross ratios straightforwardly.} $t=0,|(\vec{x}-\vec{x}_0)|=R$ and $t=0,|(\vec{x}-\vec{x}'_0)|=R'$.  We will assume  the two defects are disjoint. One can use conformal symmetry to fix $\vec{x}_0=(0,\cdots,0)$ and $\vec{x}_0'=(1,0,\cdots,0)$ and $R'=R$.  This configuration is shown in the Figure \ref{causaltwos}.
The twist operators are inserted at 
\be
X=(R,0,\cdots,0), Y=(-R,0,\cdots,0),X'=(R,1,0,\cdots,0),Y'=(-R,1,0,\cdots,0).\nn
\ee
Then any $k$-point correlation functions are
\be
<\mathcal{O}_1(x_1)\cdots\mathcal{O}_k(x_k)>_{\mathcal{D}}=\frac{<\mathcal{O}_1(x_1)\cdots\mathcal{O}_k(x_k)\tau(X)\tau(Y)\tau(X')\tau(Y')>}{<\tau(X)\tau(Y)\tau(X')\tau(Y')>}.
\ee
Note the factor in the denominator is the partition function in the presence of defects. When we discuss R\'enyi entropy, it is related to R\'enyi entropy by 
\be
S_n=\frac{1}{1-n}\log <\tau(X)\tau(Y)\tau(X')\tau(Y')>.
\ee

\section{Operator product expansion of defect operator}
As we have shown, defect operator $\mathcal{D}$ is actually a pair of twist operators.  So in principle it can be expanded as primary operators and their descendants. 
Assume all possible (quasi-)primary operators in the theory are  collected in $\{O_{\Delta,J}\}$. Then $\mathcal{D}$ can be expanded as 
\be
\mathcal{D}=Z(\mathcal{D})\sum_{\Delta,J} c_{\Delta,J}(O_{\Delta,J}+desendents).
\ee

The function $Z(\mathcal{D})$ is the expectation value of the defect,
\be
Z(\mathcal{D})=<\mathcal{D}>.
\ee
It is also the patition function of the defect CFT. 
 The `desendents' terms are fixed completely. Hence we can write the defect operators compactly as
\be
\mathcal{D}=Z(\mathcal{D})\sum_{\Delta,J} c_{\Delta,J}Q[O_{\Delta,J}]\label{ope}.
\ee
We normalize the coefficient before identity operator to be 1.  All other coefficients $c_{\Delta,J}$ are fixed by one point function of its corresponding operator $\mathcal{O}_{\Delta,J}$ hence it should be propotional to $a_{\Delta,J}$. 
The operator $Q(\mathcal{O}_{\Delta,J})$ can be fixed by shadow formalism\cite{Ferrara:19721,Ferrara:19722,Ferrara:19723,Ferrara:19724,Duffin:1204}. For a flat defect operator which sits at $t=0, x^1=0$, the operator $Q(\mathcal{O})$ is\footnote{Interestingly, similar operators also appear in other context\cite{Boer:1509,Czech:1604, Boer:1606}}
\be
Q(\mathcal{O}_{\Delta,J})=\int_{D(x^1<0)} d^d\xi K^{\mu_1}\cdots K^{\mu_J}|K|^{\Delta-J-d}O_{\mu_1\cdots\mu_J}.
\ee
Here $\mathcal{O}_{\mu_1\cdots\mu_J}$ is a general symmetric traceless operator\footnote{There may be operators with mixed symmetry in the theory. We ignore them in this paper. It would be interesting to include them into the story as they always appear in a general theory.} with dimension $\Delta$ and spin J.  Vector $K$ is 
\be
K=x^1\partial_t+t\partial_{x^1}. \label{modu}
\ee
Interestingly it is the modular Hamiltonian generator in the field of entanglement entropy.  The integral region is over the causal diamond of flat defect.  We can fix the relation between $c_{\Delta,J}$ and $a_{\Delta,J}$ by one point correlation functions \footnote{Interesting reader can find the case of spin 1 and spin 2 in Appendix D.} 
\bea
<O_{\Delta}>_{\mathcal{D}}&=&c_{\Delta}\int_{D(x^1<0)}d^d\xi(x'^2-t'^2)^{\frac{\Delta-d}{2}}<\mathcal{O}(x)\mathcal{O}(\xi)>\nn\\
&=&c_{\Delta}N_{\Delta}\frac{\pi^{\frac{d-2}{2}}\Gamma[\Delta-\frac{d-2}{2}]}{2\Gamma[\Delta]}\int_0^{\infty}du'dv'\frac{(u'v')^{\frac{\Delta-d}{2}}}{((u-u')(v-v'))^{\Delta-\frac{d-2}{2}}}\nn\\&=&
c_{\Delta}N_{\Delta}\frac{\pi^{\frac{d-2}{2}}\Gamma[\Delta-\frac{d-2}{2}]}{2\Gamma[\Delta]}|\frac{\Gamma[\frac{\Delta}{2}]\Gamma[\frac{\Delta-d+2}{2}]}{\Gamma[\Delta-\frac{d}{2}+1]}|^2\frac{1}{|x|^{\Delta}}\label{opt2}
\eea
where we have defined\footnote{Don't be confused with the cross ratio defined in the previous section.}$u=x+t,v=x-t,|x|^2=u v$. $N_{\Delta}$ is the normalization of the two point function of the operator $\mathcal{O}_{\Delta}$. Comparing (\ref{scalar})and (\ref{opt2}), we find the identification of $c_{\Delta}$ and $a_{\Delta}$ up to some constant coefficient and normalization factor
\be
a_{\Delta,0}=c_{\Delta}N_{\Delta}\frac{\pi^{\frac{d-2}{2}}}{2\Gamma[\Delta]}\frac{\Gamma[\frac{\Delta}{2}]^2\Gamma[\frac{\Delta-d+2}{2}]^2}{\Gamma[\Delta-\frac{d}{2}+1]}.\label{ac}
\ee
At first glance, we meet a problem.  For a unitary conformal field theory, the conformal dimension of a scalar primary operator has a lower bound\cite{Marc:1977},
\be
\Delta\ge\frac{d-2}{2}.
\ee
That means for some operators with dimension $\frac{d-2}{2}\le\Delta\le d-2$, $a_{\Delta,0}$ or $c_{\Delta,0}$ should be ill defined since $\Gamma(x)$ function is divergent when $x$ is a non positive integer. 
We can cure this problem by absorbing possible divergence into the definition of $Q(\mathcal{O})$.
We also checked that the coefficient $c_{\Delta,0}$ is the same for flat and spherical defect.  Note for spherical defect, the vector $K$ is not 
(\ref{modu}), its form can be found in Appendix E. Again, it is the modular Hamiltonian generator of spherical region in entanglement entropy.
\section{Applications}
In this section we try to compute mutual R\'enyi entropy of two spherical region for a free scalar theory and holographic CFTs. 
\subsection{Setup}
The system is in vacuum $|0>$, its density matrix is $\rho=|0><0|$. We choose a spatial region $A$ at a constant time slice,
its complement is denoted as $A^c$. Then one can integrate out the degree of freedom in region $A^c$ to find a reduced density matrix $\rho_A$, 
\be
\rho_A=tr_A\rho
\ee
Renyi entropy is defined as 
\be
S^{(n)}_A=\frac{\log tr\rho^n_A}{1-n}.
\ee
Entanglement entropy between $A$ and $A^c$ is the $n\to1$ limit of R\'enyi entropy, 
\be
S_A^{EE}=\lim_{n\to1}S^{(n)}_A.
\ee
R\'enyi entropy(and entanglement entropy) satisfy area law, 
\be
S^{(n)}_A=\gamma \frac{Area(\partial A)}{\epsilon^{d-2}}+\cdots
\ee
where $\epsilon$ is UV cutoff of the theory. The coefficient $\gamma$ is not universal. Universal information are encoded in the logrithmic or constant terms. For example, in any two dimensional CFTs, one interval R\'enyi entropy is 
\be
S^{(n)}_{\Delta x}=\frac{c}{6}(1+\frac{1}{n})\log\frac{|\Delta x|}{\epsilon},
\ee
where $\Delta x$ is the length of the interval, $c$ is the central charge of the CFT which is invariant under rescaling of the cutoff.  In four dimensional CFT, the universal terms of Renyi entropy are expected to be\cite{Fursaev:1201}
\be
S^{(n)}_A=\cdots+(-\frac{f_a(n)}{2\pi}\int_{\partial A}R_{\partial A}-\frac{f_b(n)}{2\pi}\int_{\partial A}(\tilde{K}_{ij}^a)^2-\frac{f_c(n)}{2\pi}\int_{\partial A}C^{ab}_{\ ab})\log R/\epsilon, \label{renyi4}
\ee
Where $R_{\partial A}$ is the Ricci scalar of the boundary of $A$, $\tilde{K}_{ij}^a$ is the traceless extrinsic curvature in the transverse direction $x^a$, $C^{ab}_{\ ab}$ is the projected Weyl tensor in the orthogonal  direction of the boundary of $A$, R is a characteristic size of region $A$. 

When $A$ is the union of  two (or more) disjoint regions, 
\be
A=A_1\cup A_2, A_1\cap A_2=\phi, 
\ee
one can define a finite quantity 
\be
I_{A_1,A_2}^{(n)}=S_{A_1}^{(n)}+S_{A_2}^{(n)}-S_{A_1\cup A_2}^{(n)}
\ee
which is called (n-th) mutual R\'enyi entropy.

Let's begin with one spherical region $A$, 
\be
A=\{t=0,\vec{x}^2=R^2\}.
\ee
As we have mentioned in previous sections, the R\'enyi entropy can be computed by inserting two twist operators at $X=(R,0,\cdots,0)$ and $Y=(-R,0,\cdots,0)$
\be
S^{(n)}_A=\frac{1}{1-n}\log<\mathcal{D}>=\frac{1}{1-n}\log <\tau(X)\tau(Y)>=\frac{1}{1-n}\log{N_{\tau}}+\frac{-2\delta}{1-n}\log 2R/\epsilon.
\ee
$N_{\tau}$ is the normalization of two point function of twist operators. In the last step we have used the fact that the dimension of the twist operator is $\delta$ and inserted a UV cutoff $\epsilon$.  For $d=2$, 
\be
\delta=\frac{c}{12}(n-\frac{1}{n}). 
\ee
This is well known in two dimensional CFTs. For $d=4$, the second and third term in (\ref{renyi4}) are zero while the first term contributes 
\be
S^{(n)}_A=-4f_a(n)\log R/\epsilon.
\ee
So the scaling dimension of the twist operator is 
\be
\delta=2(1-n)f_a(n).\label{cd}
\ee
Unfortunately, our definition of dimension of twist operator is not in agreement with \cite{Hung:1110}. In that paper, the conformal dimension of the twist operator is defined by  one point function of the stress tensor. Two definitions match in $d=2$ and doesn't match in higher dimensions. However, from the general discussion of defect operators, (\ref{cd}) may be the correct definition of the dimension of twist operator in four dimension . The conformal dimension defined in \cite{Hung:1110}  may be interpreted as the three point function coefficient of stress tensor and two twist operators. However, in even dimensions, they are actually related to each other\cite{Aitor:1407}.

In odd dimensions, there maybe no $\log R/\epsilon$ universal terms in  Renyi entropy. So we may conclude 
\be
\delta=0, \ \ for\ \ d\ \ odd.
\ee
There is a constant universal term in odd dimensions, 
\be
S_{A}^{(n)}=\cdots+q_n.
\ee
This term can be reproduced from the normalization constant $N_{\tau}$. 

\subsection{General structure of mutual R\'enyi entropy}
Let us assume region $A_1$ and $A_2$ are
\be
A_1=\{t=0,\vec{x}^2\le R^2\},A_2=\{t=0,(\vec{x}-\vec{x}_0)^2\le R'^2\}
\ee

We can use conformal symmetry to set $R'=R$ and $\vec{x}_0=(1,0,\cdots,0)$. In this case, there is only one independent cross ratio
\be
z=4R^2=\bar{z}, \ u=z\bar{z},\ v=(1-z)(1-\bar{z}).
\ee
The R\'enyi entropy is 
\be
S^{(n)}_{A_1\cup A_2}=\frac{1}{1-n}\log <\mathcal{D}_{A_1\cup A_2}>.
\ee
The expectation value of twist operator is
\be
\frac{<\mathcal{D}_{A_1\cup A_2}>}{<\mathcal{D}_{A_1}><\mathcal{D}_{A_2}>}=\sum_{\Delta,J}c_{\Delta,J}^2<Q^{A_1}_{\Delta,J}Q^{A_2}_{\Delta,J}>\sim\sum_{\Delta,J}\frac{a_{\Delta,J}^2}{N_{\Delta,J}}G_{\Delta,J}(u,v).
\ee
The building block of mutual Renyi entropy would be the following block
\be
<Q^{A_1}_{\Delta,J}Q^{A_2}_{\Delta,J}>.
\ee
We already know the answer should be a conformal partial wave with dimension $\Delta$ and spin J
\be
<Q^{A_1}_{\Delta,J}Q^{A_2}_{\Delta,J}>\sim G_{\Delta,J}(u,v).\label{cb}
\ee
What is left is to fix the total coefficient appeared before (\ref{cb}).  

The contribution of a scalar operator with dimension $\Delta$ is
\bea
<Q^{A_1}_{\Delta,J=0}Q^{A_2}_{\Delta,J=0}>&=&N_{\Delta,0}(\frac{R}{2})^{2\Delta}\int d\xi d\bar{\xi}d\xi'd\bar{\xi}'d\Omega_{d-2}d\Omega_{d-2}'|\xi+\bar{\xi}|^{d-2}|\xi'+\bar{\xi}'|^{d-2}\times\nn\\&&
\frac{((1-\xi^2)(1-\bar{\xi}^2)(1-\xi'^2)(1-\bar{\xi}'^2))^{\frac{\Delta-d}{2}}}{(R\frac{\xi+\bar{\xi}}{2}\vec{\omega}-R\frac{\xi'+\bar{\xi}'}{2}\vec{\omega}'-\vec{x}_0)^2-(\frac{\xi-\bar{\xi}}{2}-\frac{\xi'-\bar{\xi}'}{2})^2R^2)^{\Delta}}
\eea
It should be a conformal partial wave of a scalar operator in d dimension. We choose $d=4$ as an example. In this case,
\be
<Q^A_{\Delta,J=0}Q^B_{\Delta,J=0}>=N_{\Delta,0}\frac{\pi^4 2^{4-4\Delta}\Gamma[\frac{\Delta}{2}-1]^4}{\Gamma[\frac{\Delta-1}{2}]^2\Gamma[\frac{\Delta+1}{2}]^2}G_{\Delta,0}(u,v).
\ee
Note when $z=\bar{z}$, the conformal partial wave is\footnote{This can be obtained by taking $\bar{z}\to z$ limit of general conformal partial waves.}
\bea
G_{\Delta,J}(u,v)&=&(-1)^J\frac{z^{\Delta}}{4}(z(2+J-\Delta) {}_2F_{1}(\frac{\Delta-J}{2},\frac{\Delta-J}{2},\Delta-J-1,z)\nn\\&&\times {}_2F_{1}(\frac{\Delta+J}{2},\frac{\Delta+J}{2},\Delta+J,z)\nn\\
&&+4(J+1) {}_2F_{1}(\frac{\Delta-J-2}{2},\frac{\Delta-J-2}{2},\Delta-J-2,z){}_2F_{1}(\frac{\Delta+J}{2},\frac{\Delta+J}{2},\Delta+J,z)\nn\\&&+z(\Delta+J){}_2F_{1}(\frac{\Delta-J-2}{2},\frac{\Delta-J-2}{2},\Delta-J-2,z)\nn\\&&\times {}_2F_{1}(\frac{\Delta+J+2}{2},\frac{\Delta+J+2}{2},\Delta+J+1,z)).\label{conformal}
\eea
Combining the relation between  $c_{\Delta,0}$ and $a_{\Delta,0}$, we find
\be
c_{\Delta,0}^2<Q^{A_1}_{\Delta,0}Q^{A_2}_{\Delta,0}>=\frac{a_{\Delta,0}^2}{N_{\Delta,0}}G_{\Delta,0}(u,v).
\ee
In the same way, we find 
\be
c_{\Delta,1}^2<Q^{A_1}_{\Delta,1}Q^{A_2}_{\Delta,1}>=\frac{a_{\Delta,1}^2}{2N_{\Delta,1}}G_{\Delta,1}(u,v), 
\ee
\be
c_{\Delta,2}^2<Q^{A_1}_{\Delta,J=2}Q^{A_2}_{\Delta,J=2}>=\frac{4 a_{\Delta,2}^2}{N_{\Delta,2}}G_{\Delta,2}(u,v).
\ee
Hence the mutual R\'enyi entropy of two spherical region is 
\be
I^{(n)}_{A_1,A_2}=-\frac{1}{1-n}\log (1+\sum_{\Delta}(s_{\Delta,0}G_{\Delta,0}+s_{\Delta,1}G_{\Delta,1}+s_{\Delta,2}G_{\Delta,2}+\cdots)).
\ee
where $s_{\Delta,0}=\sum_{\mathcal{O}}\frac{a^2_{\mathcal{O}}}{N_{\mathcal{O}}}$ is to sum over the contributions of all possible scalar operators with dimension $\Delta$. The generalization to other dimensions is straightforward. In our case, $\bar{z}=z$, the conformal partial waves can be found in \cite{Matthijs:1305}  for general dimensions. 

\subsection{Free scalar theory}
Let us consider the lowest dimension operators at first. They are 
\be
\mathcal{O}=\phi^2,\ \mathcal{O}_{ij}=\phi_i\phi_j (i\not=j).
\ee
We assume the normalization of two point function is 1,
\be
<\phi(x)\phi(y)>=\frac{1}{(x-y)^{d-2}}.
\ee
The position $x,x'$ can be\footnote{r is radial direction in the orthogonal plane and $\theta$ is angular direction. We have switched to Euclidian signature in this section.}
\be
x=(r,\theta,y^i),x'=(r',0,\cdots,0).
\ee
In $d=4$,  one can find\cite{Guimaraes:1994, Nozaki:1401}
\be
<\phi(x)\phi(x')>_n=\frac{\sinh\frac{\eta}{n}}{2nrr'\sinh\eta(\cosh\frac{\eta}{n}-\cos\frac{\theta}{n})}.\label{tptfree}
\ee
The function $\eta$ is defined as 
\be
\cosh\eta=\frac{r^2+r'^2+y^2}{2rr'}.
\ee

By symmetry, the one point function of $\mathcal{O}$ and $\mathcal{O}_{ij}$ should be 
\be
<\phi^2>_n=\frac{a_{\phi^2}}{r^{2}},\ <\phi_i\phi_j>_n=\frac{a_{ij}}{r^{2}}.
\ee
To compute the coefficients $a_{\phi^2}$ and $a_{ij}$, one can set $y=0$ and $r'=r$ in (\ref{tptfree}) and read out small $\theta$ limit terms.
 The coefficients $a_{\mathcal{O}}$ and $a_{ij}$ 
\be
a=\frac{1-n^2}{12n^2},a_{ij}=\frac{1}{4n^2\sin^2\frac{\theta_{ij}}{2n}} 
\ee
where $\theta_{ij}=2\pi(i-j)$. 
The normalization factor of $\phi^2$ and $\phi_i\phi_j$ are 
\be
N_{\phi^2}=2,\ N_{\phi_i\phi_j}=1.
\ee
Then 
\be
s_{\Delta=2,J=0}=\frac{n a_{\phi^2}^2}{2}+\sum_{j_1>j_2}a_{j_1j_2}^2=\frac{n^4-1}{240n^3}.
\ee
 Mutual R\'enyi entropy from lowest dimension operator 
\be
I^{(n)}_{A_1,A_2}=\frac{1}{n-1}\log(1+\frac{n^4-1}{240n^3}G_{\Delta=2,J=0}(u,v)+\cdots).
\ee
$G_{\Delta,J}$ is defined in (\ref{conformal}). Note when the two region $A_1$ and $A_2$ are far away to each other, $z\to0$, then the leading contribution is from lowest dimension operator, in this limit 
\be
I^{(n)}_{A_1,A_2}=\frac{(n+1)(n^2+1)}{240n^3}z^2+\cdots
\ee
This is exactly those found in \cite{Shiba:1207, Cardy:1304}. At next order, we must consider the spin 1 operator 
\be
J^{(ij)}_{\mu}=\phi_i\partial_{\mu}\phi_j-(i\leftrightarrow j),\ i>j
\ee
We can also read out the expectation value of $J^{(ij)}_{\mu}$, 
\be
<J^{(ij)}_a>_n=a_J^{(ij)}\frac{\epsilon_{ab}x^b}{|x|^4},\ <J^{(ij)}_k>_n=0.
\ee
where 
\be
a_J^{(ij)}=\frac{\cos\frac{\theta_{ij}}{2n}}{2n^3\sin^3\frac{\theta_{ij}}{2n}}.
\ee
Normalization factor
\be
N_{J}=4
\ee
can be read from two point function 
\be
<J_{\mu}^{(ij)}(x)J_{\nu}^{(ij)}(x')>=4\frac{\eta_{\mu\nu}-2\frac{(x-x')_{\mu}(x-x.)_{\nu}}{|x-x'|^2}}{|x-x'|^6}.
\ee
So
\be
s_{\Delta=3,J=1}=\sum_{i>j}\frac{(a_J^{(ij)})^2}{2N_J}=\frac{n^6-21n^2+20}{30240n^5}.
\ee
After including spin 1 operators, the mutual R\'enyi entropy becomes
\be
I^{(n)}_{A_1,A_2}=\frac{1}{n-1}\log(1+\frac{n^4-1}{240n^3}G_{\Delta=2,J=0}(u,v)+\frac{n^6-21n^2+20}{30240n^5}G_{\Delta=3,J=1}(u,v)+\cdots)
\ee
In the limit of small cross ratio,  $z^2$ and $ z^3$ terms are exact. 


\subsection{Holographic R\'enyi entropy}
In a holographic CFT which is dual to Einstein gravity, the operators appear in the dual CFT is just the stress tensor and its composites. The lowest dimension operator appear in the theory is the stress tensor whose one point function is
\be
<T_{ab}>_n=\frac{h_n}{2\pi}\frac{(d-1)\delta_{ab}-d\frac{x_ax_b}{x^2}}{|x|^d},\ <T_{ij}>_n=-\frac{h_n}{2\pi}\frac{\delta_{ij}}{|x|^d}
\ee 
For Einstein gravity, we have\footnote{Here $T_{\mu\nu}$ is a single sheet stress tensor.}\cite{Hung:1110}
\be
h_n=\frac{1}{8G}x_n^{d-2}(1-x_n^2),\ x_n=\frac{1}{nd}(1+\sqrt{1-2dn^2+d^2n^2}),\ N_{T}=C_{T}=\frac{1}{8\pi G}\frac{\Gamma[d+2]}{\pi^{\frac{d}{2}}(d-1)\Gamma[\frac{d}{2}]}.
\ee
Now in $d=4$, we get the mutual R\'enyi entropy
\be
I^{(n)}_{A_1,A_2}=\frac{1}{n-1}\log(1+n\times \frac{4(-\frac{h_n}{2\pi})^2}{N_T}G_{\Delta=4,J=2}+\cdots).
\ee
In the small cross ratio limit, we can expand it in terms of small z,
\be
I^{(n)}_{A_1,A_2}=\frac{n}{n-1}\frac{C_T \pi^4 x_n^4(1-x_n^2)^2}{1600}(3z^4+6z^5+\frac{60z^6}{7}+\frac{75z^7}{7})+\mathcal{O}(z^8).
\ee
Note at $\mathcal{O}(z^8)$, there may contribute from $T^2$, so mutual R\'enyi entropy will be modified from this order. 

More interesting, if we assume all other operators don't contribute at order $\mathcal{O}(n-1)$
\bea
I^{(n)}_{A_1,A_2}&=&\frac{\pi^4 C_T(n-1)}{3600}G_{\Delta=4,J=2}+\mathcal{O}(n-1)^2\nn\\&=&\frac{\pi^4C_T(n-1)}{120}(11+\frac{1}{1-z}-z-\frac{6(z-2)\log(1-z)}{z})+\mathcal{O}(n-1)^2.
\eea
This agrees with \cite{Dong:1601}. We checked that the incomplete Beta function is indeed a conformal partial wave with $\Delta=4,J=2$. 

For general d, if we normalize the behaviour of the conformal partial wave\footnote{There is a normalization factor 3 compare to $d=4$ conformal partial wave we used above.}
\be
G^{(d)}_{\Delta,2}=z^{\Delta}+\cdots
\ee
in the small $z$ limit, then we find 
\be
I^{(n)}_{A_1,A_2}=\frac{1}{n-1}\log (1+d(d-1)n \frac{a_T^2}{N_T}G^{(d)}_{\Delta=d,J=2}+\cdots)
\ee

The result is exact up to order $z^{2d}$. 
Again,
\be
I^{(n)}_{A_1,A_2}=\frac{C_T d(d-1)\pi^d\Gamma[\frac{d}{2}]^2}{\Gamma[d+2]^2}(n-1)G^{(d)}_{\Delta=d,J=2}(z)+\mathcal{O}(n-1)^2
\ee
It agrees with computation from gravity side\cite{Dong:1601}.

\section{Conclusion and discussion}
In this work, we studied flat or spherical co-dimension two defect operator $\mathcal{D}$. We conjecture that it is a pair of twist operators which are inserted at the tips of its causal diamond. In this way, we can map any $k$-point correlation function in defect CFT to a $(k+2)$-correlation function.  Especially, one point correlation function of a primary operator is completely fixed up to a constant which can be interpreted as a three point correlation function coefficient of the operator and two twist operators. We reproduce known even spin results of previous paper. We also find one point correlation function of odd spin . The spin 1 case matches with previous work. As far as we know, odd higher spin results are new.  Interestingly, parity violating term appears naturally in this method.  The parity violating term may appear even though the original CFT is parity invariant as we have shown for free scalar theory. When we apply it to R\'enyi entropy, we give a description of scaling dimension to twist operator. In our description, scaling dimension of the twist operator is related to the universal terms of R\'enyi entropy in even dimensions. In odd dimensions, the scaling dimension of the twist operator is zero.  The conformal dimension defined in \cite{Hung:1110} can be naturally interpreted as the three point function coefficient between stress tensor and two twist operators.

Two point functions of defect CFT are more intriguing. They can be mapped to four point correlation functions.  So it can be expanded in terms of conformal partial waves. In this way, we quickly read out the solution of the differential equation which is imposed by symmetry constraints.  The same method goes through to other tensor fields.  

When there are multi-defects, we should insert a pair of twist operators for each defect. We apply this idea to compute mutual R\'enyi entropy of two spherical region.  Interestingly, this mutual R\'enyi entropy is essentially four point function of twist operators. So it can also be expanded as conformal partial waves. One just need to fix the one point function for each primary operator in the defect CFT. For 4d free scalar theory, we reproduce the result of \cite{Cardy:1304}  in the small cross ratio limit. We also obtain subleading terms by including the contribution of spin 1 operators.  For holographic CFTs, we reproduce the result of \cite{Dong:1601} in $(n-1)$ expansion in general dimensions.  In the limit of small cross ratios, the leading d terms are exact. These results strongly support the idea of inserting twist operators. It provides a unified way to compute multi spherical region R\'enyi entropy in arbitrary dimensions.  

In the following, we will discuss some problems which may be further explored.
\begin{itemize}
\item Non-universal and universal terms. 
We have shown that for $k>0$, the correlation functions can be maped to $(k+2)$-point correlation functions. For $k=0$, one should expect
\be
\log Z(\mathcal{D})=\log <\tau(X)\tau(Y)>.\label{0pt}
\ee
To apply it to R\'enyi entropy, we split the R\'enyi entropy by universal and non-universal terms in even dimension, then 
\be
S_{A}^{(n)}=S_{non-uni}+s_{uni}\log R/\epsilon.\label{ry}
\ee
To match (\ref{0pt}) and (\ref{ry}), we find 
\be
N_{\tau}=e^{(1-n)S_{non-uni}},\ \delta=-\frac{(1-n)s_{uni}}{2}.
\ee
Area terms(and non-universal subleading terms) can be absorbed into the normalization of two point function of twist operator.  In odd dimensions, we have 
\be
S_{A}^{(n)}=S_{non-uni}+q_n, 
\ee
where $q_n$ is universal which is invariant under scaling of the cutoff. In this case, 
\be
N_{\tau}=e^{(1-n)S_A^{(n)}},\ \delta=0.
\ee
From the point of view of twist operator, the appearance of the universal term is inevitable once the scaling dimension of the twist operator is non-zero. When the scaling dimension of the twist operator is zero, there may be universal terms which is encoded in the normalization factor $N_{\tau}$.  We will conjecture there may be universal terms in general defect CFTs. It would be nice to check this point in other explicit examples. 
\item Two dimensions. We noticed that in $d=2$, the twist operators are inserted at the boundary of the interval. In our case, the twist operator should be inserted at the tips of the causal diamond of the interval. We argue that there is no contradiction between the two different descriptions.  To see it,  the causal diamond is actually the same for the two descriptions. 
\item Defect channel partial waves. The bulk channel correlation functions can be obtained by inserting twist operators. However, in \cite{Marco:1601},  two bulk operator correlation functions can also be expanded in terms of defect channel partial waves. There will be infinite number of equalities for three point function coefficients between different channels. It seems to be possible to fix all coefficients of defect channel in terms of bulk channel information.
\item Displacement operator. We only consider flat or spherical defect operator in this work,  a general co-dimension two defect operator can be obtained by inserting displacement operator. Some results have been obtained \cite{Srivatsan:1607, Bianchi:1607}. It would be interesting to seek how to define twist operators in general situations.
\end{itemize}
\section*{Acknowledgments}

The author would like to thank Bin Chen, Geoffrey Comp\`{e}re and Marco Meineri for their comments on the manuscript. This work is supported by the ERC Starting Grant 335146 ``HoloBHC".

\appendix
\section{Conventions}
In this paper, $\mu,\nu,\cdots$ are spacetime coordinates. For co-dimension two defects, we split the coordinate to the direction which is parallel to the defect and orthogonal to the defect. $i,j,\cdots$ are the direction which is parallel to the defect. $a,b,\cdots$ are the direction which is orthogonal to the defect. In the same way, the metric $\eta_{\mu\nu}$ splits to 
$\eta_{ab}$ and $\eta_{ij}$. When we need to distinguish time and spatial direction, t is time direction and $\vec{x}$ are the $d-1$ spatial direction. In the embedding space formalism, the coordinate index is labeled by capital $M,N,\cdots$. 
\section{Embedding space formalism}
This formalism is especially powerful when we study tensor like operators. In this formalism, we map any point x in $CFT_d$ to a light cone of $d+2$ dimensional Minkowski spacetime.  A point $P$ in the Poincare section is 
\be
P=(1,x^2,x^{\mu}).
\ee
We use the light cone coordinates, in other words
\be
P^2=-P^+P^-+\eta_{\mu\nu}P^2.
\ee
More generally, we define a scalar product as 
\be
P_1\cdot P_2=P_1^MP_2^N \eta_{MN}.
\ee
A primary symmetric traceless tensor  $\mathcal{O}_{\mu_1\cdots\mu_J}$  in the original $CFT$ is mapped to a symmetric traceless and transverse tensor $\mathcal{O}_{M_1\cdots M_J}$ with degree $-\Delta$, 

\bea
\mathcal{O}_{\mu_1\cdots\mu_J}&=&\pi^{M_1\cdots M_J}_{\mu_1\cdots\mu_J}\mathcal{O}_{M_1\cdots M_J}, \nn\\
P^{M_1}\mathcal{O}_{M_1\cdots M_J}&=&0,\\
\mathcal{O}_{M_1\cdots M_J}(\lambda P)&=&\lambda^{-\Delta}\mathcal{O}_{M_1\cdots M_J}(P).
\eea
The projector 
\be
\pi^{M_1\cdots M_J}_{\mu_1\cdots\mu_J}=\frac{\partial P^{M_1}}{\partial x^{\mu_1}}\cdots \frac{\partial P^{M_J}}{\partial x^{\mu_J}}.
\ee
To avoid the index structure of the operator, we uplift it to a polynomial of Z,
\be
\mathcal{O}_{\Delta,J}(P,Z)=\mathcal{O}_{M_1\cdots M_J}(P)Z^{M_1}\cdots Z^{M_J},   Z^2=0.
\ee
For a flat defect which sits at $t=0,x^1=0$, we may split the index 
\be
M=(A,I)
\ee
where $A=0,1$ and $I=+,-,2,\cdots,d-1$, the metric $\eta_{MN}$ split to $\eta_{AB}$ and $\eta_{IJ}$.  Two scalar quantities can be defined as 
\be
P_1\bullet P_2=\eta_{IJ}P_1^IP_2^J,\ P_1\circ P_2=\eta_{AB}P_1^AP_2^B.
\ee
We will also embed the twist operators into $d+2$ dimensions.  They are inserted at 
\be
X=(1,0,T,-T,\cdots),\ Y=(1,0,-T,-T,\cdots). 
\ee
The $\cdots$ terms are not important as $T\to\infty$ in the end. Three point function of an spin J with two other scalar operator is 
\be
<\mathcal{O}_1(P_1)\mathcal{O}_2(P_2)\mathcal{O}_{\Delta,J}(P,Z)>=\frac{c_{12,\{\Delta,J\}}}{(P_1\cdot P_2)^{\frac{\Delta_1+\Delta_2-\Delta-J}{2}}(P_2\cdot P)^{\frac{\Delta_2+\Delta-J-\Delta_1}{2}}(P\cdot P_1)^{\frac{\Delta+\Delta_1-\Delta_2-J}{2}}}\mathcal{X}_{12}(P,Z)^J
\ee
where 
\be
\mathcal{X}_{12}(P,Z)=\frac{P_1\cdot Z}{P_1\cdot P}-\frac{P_2\cdot Z}{P_2\cdot P}. 
\ee
According our conjecture, 
after some simple algebra, we find 
\be
<\mathcal{O}_{\Delta,J}(P,Z)>_{\mathcal{D}}=a_{\Delta,J}\frac{(\epsilon_{AB}Z^AP^B)^J}{(P\circ P)^{\frac{\Delta+J}{2}}}. 
\ee
where we redefine the total coefficient to be $a_{\Delta,J}$. We also defined an epsilon tensor with $\epsilon_{01}=1$. The appearance of $\circ$ product and $\epsilon_{AB}$ is due to the fact that only the orthogonal directions are relevant in the $T\to\infty$ limit. Note for even spins, we can contract two $\epsilon$ tensors to find the same answer as\cite{Marco:1601}. 

\section{Differential operators}
The differential operator in (\ref{ca1}) is 
\bea
D_{bulk}&=&2\xi^2(2+\xi\cos\phi+2\cos^2\phi)\frac{\partial^2}{\partial\xi^2}+2\sin^2\phi(2\sin^2\phi-\xi\cos\phi)\frac{\partial^2}{\partial\cos\phi^2}\nn\\
&&-4\xi\sin^2\phi(\xi+2\cos\phi)\frac{\partial^2}{\partial\xi\partial\cos\phi}+2\xi(2(1+\cos^2\phi)-(2d-\xi\cos\phi))\frac{\partial}{\partial\xi}\\
&&(2\xi\cos^2\phi-4\cos\phi\sin^2\phi)\frac{\partial}{\partial\cos\phi}-(\Delta_{12}^2\cos\phi(\cos\phi+\frac{\xi}{2})-\Delta_{12}^2+2C_{\Delta,J}),\nn
\eea
where $\Delta_{12}=\Delta_1-\Delta_2$, $C_{\Delta,J}=\Delta(\Delta-d)+J(J+d-2)$ is  eigenvalue of $so(d,2)$ Casimir operator.  The differential operator in (\ref{ca2}) is 
\bea
D_{CFT}&=&-2((1-v)^2-u(1+v))\frac{\partial}{\partial v}v\frac{\partial}{\partial v}-2u(1-u+v)\frac{\partial}{\partial u}u\frac{\partial}{\partial u}+4uv(1+u-v)\frac{\partial^2}{\partial u\partial v}\nn\\
&&+2du\frac{\partial}{\partial u}-\Delta_{12}((1+u-v)(u\frac{\partial}{\partial u}+v\frac{\partial}{\partial v})-(1-u-v)\frac{\partial}{\partial v}+C_{\Delta,J}.
\eea
One can directly check the following equivalence
\be
D_{bulk}F_{\Delta,J}=0\leftrightarrow D_{CFT}G_{\Delta,J}=0
\ee
when using (\ref{cross}) and (\ref{fg}). 
\section{Relation between $c_{\Delta,J}$ and $a_{\Delta,J}$}
In section 4 we constructed the operator product expansion of defect operator $\mathcal{D}$. We have mentioned that the normalization constant $c_{\Delta,J}$ is proportional to $a_{\Delta,J}$.  In this appendix, we will determine the proportional constant for spin 1 and spin 2. 
\bea
<\mathcal{O}^{a}>_{\mathcal{D}}&=&c_{\Delta,1}\int_{D(x^1<0)}d^d\xi|K'|^{\Delta-d-1}K'^{\nu}<\mathcal{O}^a(x)\mathcal{O}_{\nu}(x')>\nn\\&=&c_{\Delta,1}N_{\Delta,1}\frac{\pi^{\frac{d-2}{2}}\Gamma[\frac{\Delta+1}{2}]^2\Gamma[\frac{\Delta-d+1}{2}]\Gamma[\frac{\Delta-d+3}{2}]}{\Gamma[1+\Delta]\Gamma[\Delta-\frac{d}{2}+1]}\frac{\epsilon^{ab}x_b}{|x|^{\Delta+1}},\\
<\mathcal{O}^{i}>_{\mathcal{D}}&=&0.
\eea
Then we find 
\be
a_{\Delta,1}=c_{\Delta,1}N_{\Delta,1}\frac{\pi^{\frac{d-2}{2}}\Gamma[\frac{\Delta+1}{2}]^2\Gamma[\frac{\Delta-d+1}{2}]\Gamma[\frac{\Delta-d+3}{2}]}{\Gamma[1+\Delta]\Gamma[\Delta-\frac{d}{2}+1]}
\ee
for spin 1 operator. To study spin 2 operator, we can just compute the components with parallel direction   
as all other components are completely fixed by it. 
\be
<\mathcal{O}_{ij}>_{\mathcal{D}}=c_{\Delta,2}N_{\Delta,2}\frac{\pi^{\frac{d-2}{2}}\Delta(\Delta-d+1)\Gamma[1+\frac{\Delta}{2}]\Gamma[\frac{\Delta}{2}]\Gamma[\frac{\Delta-d}{2}]\Gamma[\frac{\Delta-d+2}{2}]}{2d\Gamma[2+\Delta]\Gamma[\Delta-\frac{d}{2}+1]}\frac{\eta_{ij}}{\rho^{\Delta}},
\ee 
we read
\be
a_{\Delta,2}=c_{\Delta,2}N_{\Delta,2}\frac{\pi^{\frac{d-2}{2}}\Delta(\Delta-d+1)\Gamma[1+\frac{\Delta}{2}]\Gamma[\frac{\Delta}{2}]\Gamma[\frac{\Delta-d}{2}]\Gamma[\frac{\Delta-d+2}{2}]}{2d\Gamma[2+\Delta]\Gamma[\Delta-\frac{d}{2}+1]}.
\ee
Note for conserved current, we have
\be
\Delta=d-1
\ee
for spin 1 and 
\be
\Delta=d
\ee
for spin 2, we should redefine the charge to absorb the divergent $\Gamma$ function.

\section{Generator of Modular Hamiltonian}
When the region $A$ is half space, the modular Hamiltonian is generated by
\be
K=(x,t,0,\cdots,0).
\ee
Then
\be
K^2=-x^2+t^2.
\ee
When the region $A$ is a sphere with radius R whose center is origin, the modular Hamiltonian is
\be
K=\frac{1}{2R}(R^2-\vec{x}^2-t^2,-2tx^i).
\ee
After coordinate transformation
\be
t=\frac{\zeta-\bar{\zeta}}{2}R, x^i=\frac{\zeta+\bar{\zeta}}{2}R\omega^i
\ee
the metric becomes
\be
ds^2=R^2(d\zeta d\bar{\zeta}+\frac{(\zeta+\bar{\zeta})^2}{4}d\vec{\omega}^2).
\ee
The integration measure is\footnote{We have inserted a factor $\frac{1}{2}$ in the measure as the coordinates will cover the causal diamond of $A$ twice.}
\be
d^dx=(\frac{R}{2})^d|\zeta+\bar{\zeta}|^{d-2}d\Omega_{d-2}d\zeta d\bar{\zeta}
\ee
We find 
\be
K^2=-\frac{R^2}{4}(1-\zeta^2)(1-\bar{\zeta}^2).
\ee

\end{document}